\documentclass[pra,10pt,aps,twocolumn,floatfix,
               showpacs,noshowpacs,
               superscriptaddress, 
]{revtex4-2}
\usepackage{amsmath}
\usepackage{amssymb}
\usepackage{epsfig}
\usepackage{multirow}
\usepackage{bm}
\usepackage{color}
\usepackage{hyperref}
\usepackage{url}

\usepackage{enumitem}

\usepackage[version=3]{mhchem} 
\usepackage{siunitx}

\newcommand{\beq}{\begin{equation}}
\newcommand{\eeq}{\end{equation}}
\newcommand{\bea}{\begin{eqnarray}}
\newcommand{\eea}{\end{eqnarray}}

\newcommand{\eF}{\varepsilon_{F}}

\newcommand{\kF}{k_{\textrm{F}}}

\newcommand{\I}{\mathrm{i}}
\newcommand{\op}[1]{\hat{#1}}                     
\newcommand{\vect}[1]{\bm{#1}}                    
\newcommand{\abs}[1]{\lvert{#1}\rvert}            

\newcommand{\alphup}{\alpha_{\uparrow}}
\newcommand{\alphdw}{\alpha_{\downarrow}}
\newcommand{\ndw}{n_{\downarrow}}
\newcommand{\taup}{\tau_{\uparrow}}
\newcommand{\tadw}{\tau_{\downarrow}}

\newcommand{\nup}{n_{\uparrow}}

\newcommand{\ETF}{GPE}
\newcommand{\healinglength}{\xi}  
\newcommand{\xiBertsch}{\xi}
\newcommand{\dx}{\mathrm{d}x}
\providecommand{\exclude}[1]{}

\newcommand{\respch}[1]{{\color{blue} #1}}
\newcommand{\resprm}[1]{{\color{red} \sout{#1}}}
\renewcommand{\respch}[1]{#1}
\renewcommand{\resprm}[1]{}

\usepackage[normalem]{ulem}

\graphicspath{{./figures/}}

\begin{document}

\title{Rotating quantum turbulence in the unitary Fermi gas}

\author{Khalid Hossain}
\affiliation{Department of Physics and Astronomy, Washington State University, Pullman, Washington 99164, USA}
\email{mdkhalid.hossain@wsu.edu}

\author{Konrad Kobuszewski}
\affiliation{Faculty of Physics, Warsaw University of Technology, Ulica Koszykowa 75, 00-662 Warsaw, Poland}
\email{konrad.kobuszewski.dokt@pw.edu.pl}

\author{Michael McNeil Forbes}
\affiliation{Department of Physics and Astronomy, Washington State University, Pullman, WA 99164, USA}
\affiliation{Department of Physics, University of Washington, Seattle, Washington 98195--1560, USA}
\email{m.forbes@wsu.edu}

\author{Piotr Magierski}\email{piotrm@uw.edu}
\affiliation{Faculty of Physics, Warsaw University of Technology, Ulica Koszykowa 75, 00-662 Warsaw, Poland}
\affiliation{Department of Physics, University of Washington, Seattle, Washington 98195--1560, USA}

\author{Kazuyuki Sekizawa}\email{Present address: Department of Physics, School of Science, Tokyo Instutute of Technology, Tokyo 152-8551, Japan}
\affiliation{Center for Transdisciplinary Research, Institute for Research Promotion, Niigata University, Niigata 950-2181, Japan}
\affiliation{Division of Nuclear Physics, Center for Computational Sciences, University of Tsukuba, Ibaraki 305-8577, Japan}

\author{Gabriel Wlaz\l{}owski}\email{gabriel.wlazlowski@pw.edu.pl}
\affiliation{Faculty of Physics, Warsaw University of Technology, Ulica Koszykowa 75, 00-662 Warsaw, Poland}
\affiliation{Department of Physics, University of Washington, Seattle, Washington 98195--1560, USA}

\begin{abstract}
Quantized vortices carry the angular momentum in rotating superfluids, and are key to the phenomenon of quantum turbulence.
Advances in ultracold-atom technology enable quantum turbulence to be studied in regimes with both experimental and theoretical control, unlike the original contexts of superfluid helium experiments.
While much work has been performed with bosonic systems, detailed studies of fermionic quantum turbulence are nascent, despite wide applicability to other contexts such as rotating neutron stars.
In this paper, we present a large-scale study of quantum turbulence in rotating fermionic superfluids using an accurate time-dependent density functional theory called the superfluid local density approximation.
We identify two different modes of turbulent decay in the dynamical equilibration of a rotating fermionic superfluid, and contrast these results with a computationally simpler description provided by the Gross-Pitaevskii equation, which we find can qualitatively reproduce these decay mechanisms if dissipation is explicitly included.
    These results demonstrate that dissipation mechanisms intrinsic to fermionic superfluids play a key role in differentiating fermionic from bosonic turbulence, which manifests by enhanced damping of Kelvin waves.

\end{abstract}

\maketitle

\section{Introduction}
Quantized vortices are a direct manifestation of superfluidity.
In rotating systems, they relax into an Abrikosov lattice~\cite{Abrikosov}.
This lattice can be destroyed by external perturbations, leading to quantum turbulence---complex nonequilibrium flow with tangled vortices that collide and reconnect, transferring energy between length scales.
This energy cascade brings the system back towards equilibrium, giving rise to an effective dissipation at large scales despite the superfluid nature of the system.
Quantum turbulence has been observed in rotating systems with superfluid \ce{^3He}~\cite{Eltsov1,Eltsov2,Finne,Hosio} and \ce{^4He} experiments~\cite{Swanson, Walmsley}.
Here, we study this phenomenon in an ultracold atomic gas of strongly interacting fermions---the unitary Fermi gas (UFG)---where the ratio of the pairing gap to the Fermi energy $\Delta/\eF \approx 0.5$ attains the largest known value.
As the corresponding coherence length $\healinglength\exclude{ = 2\eF/\pi\kF\Delta} \approx 1/\kF$ becomes of the order of the average interparticle separation, the UFG can maintain a high density of vortex lines in a robust superfluid with a high Landau critical velocity
~\cite{GiorginiRMP}.
These features, in combination with experimental access, make the UFG an ideal platform to study quantum turbulence in a regime that cannot be accessed in superfluid helium or in cold-atom Bose-Einstein condensates (BECs)~\cite{BulgacForbesWlazlowski}.

Direct numerical studies of fermionic quantum turbulence are very challenging.
Due to the high vortex-line density, the mean intervortex distance is comparable to the size of the vortex core, precluding the use of mesoscopic vortex filament models (VFMs)~\cite{turbulence-book}.
Instead, a microscopic approach is required.
For weakly interacting bosons at $T=0$, a time-dependent nonlinear Schr\"{o}dinger equation is justified: the Gross-Pitaevskii equation (GPE) (see, e.g., Ref.~\cite{Pethick:2002}).
This is a form of orbital-free density functional theory (DFT), and generalizations have been used to model fermionic superfluids~\cite{PieriStrinati,Kim:2004,Salasnich:2008,Salasnich:2009,ForbesSharma,Simonucci1,Simonucci2}.
While lacking effects related to the Pauli exclusion principle, orbital-free simulations of condensate dynamics are computationally simple and capture many qualitative effects~\cite{ForbesSharma}.
We use a carefully tuned form  of such a generalized method
as a baseline for comparison.
We refer to this simply as the \ETF{} in the remainder of this paper.

To account for Pauli exclusion, one must currently use a fermionic DFT such as the Bogoliubov--de Gennes (BdG) or Hartree-Fock Bogoliubov (HFB).
Here we use the time-dependent asymmetric superfluid local density approximation
(TDASLDA)~\cite{LNP__2012,ASLDA-LOFF}, a fermionic DFT carefully validated 
against experiments and quantum Monte Carlo calculations at the few-percent 
level for static properties~\cite{FGG:2010,Forbes:2012,ASLDA-LOFF,LNP__2012}.
Quantitative validation at the same level for dynamics is ongoing~\cite{PRL__2014,PRAR__2015,SuppressedSolitonicCascade}, and a motivation for this paper.
The TDASLDA introduces significant differences from the \ETF{}, correctly accounting for the filling of fermionic vortex cores with a ``normal'' component~\cite{MachidaKoyama,Sensarma1,BulgacYu}, even at $T=0$.
The structure of the vortex core is also sensitive to spin-imbalance~\cite{Takahashietal,HuLiuDrummond}, which affects its size and generates reversed flow~\cite{spin-polarized-vortex}.
The spin imbalance of the system thus provides control over the amount of the normal component, and modifies the intensity of dissipative processes.
This work advances our understanding of fermionic superfluidity in the turbulent regime, with important consequences for superfluids in cold-atom experiments and neutron stars.
In particular, we investigate mechanisms of turbulent decay.
Comparing with the \ETF{} approach provides insight into the role of fermionic degrees of freedom in the decay process.


\section{Numerical Simulations}\label{sec:2}
We numerically simulate the UFG close to the zero temperature limit with two fully microscopic approaches: the TDASLDA with explicit fermionic degrees of freedom (orbitals), and the \ETF{} that models the condensate with a single wave function.
The TDASLDA allows us to also study spin-imbalanced systems with unequal numbers of spin up ($N_{\uparrow}$) and spin down ($N_{\downarrow}$) atoms.
This approach has been applied to a variety of systems, including ultracold atomic gases~\cite{Science__2011,PRL__2014,PRAR__2015,SuppressedSolitonicCascade}, atomic nuclei~\cite{PRL__CoulEx,PRL__2016,PRL_MSW_2017}, and neutron star crusts~\cite{VortexPinning}.
In  particular, we demonstrate~\cite{SuppressedSolitonicCascade} that the TDASLDA correctly captures the dynamics and evolution of topological excitations like quantized vortices by comparing with experiments.

The TDASLDA has the same structure as the commonly used BdG equations (here and below we use natural units where $m=\hbar=k_B=1$),
\begin{equation}\label{eqn:TDASLDA}
  i\frac{\partial}{\partial t} \begin{pmatrix}
    u_{n,\uparrow}(\textbf{r},t) \\
    v_{n,\downarrow}(\textbf{r},t)
  \end{pmatrix}
  = \begin{pmatrix}
    h_{\uparrow}(n_i,\nu) & \Delta(n_i,\nu) \\
    \Delta^*(n_i,\nu) & - h_\downarrow^*(n_i,\nu)
  \end{pmatrix}
  \begin{pmatrix}
    u_{n,\uparrow}(\textbf{r},t) \\
    v_{n,\downarrow}(\textbf{r},t)
  \end{pmatrix},
\end{equation}
where the single particle Hamiltonian $h_{i}$ and pairing field $\Delta$ depend on the normal $n_{i}$ and anomalous $\nu$ densities ($i=\uparrow,\downarrow$)
\begin{align}
    n_{i}(\textbf{r},t) &= \sum_{|E_n|<E_c} |v_{n,i}(\textbf{r},t)|^2 f_{\beta}(-E_n),\label{eqn:dens_n}\\
    \nu(\textbf{r},t) &= \sum_{|E_n|<E_c}  v_{n,\downarrow}^{*}(\textbf{r},t) u_{n,\uparrow}(\textbf{r},t)\frac{f_{\beta}(-E_n)-f_{\beta}(E_n)}{2}\label{eqn:dens_nu}.
\end{align}
The densities are parametrized in terms of Bogoliubov quasiparticle wave functions $\{ v_{n,i}, u_{n,i} \}$.
Note that the contribution from each quasiparticle state of energy $|E_n|$ smaller than the cutoff energy $E_c$ is weighted by the Fermi-Dirac distribution $f_{\beta}(E) = 1/[\exp(\beta E)+1]$ where $\beta=1/T$. This weighting minimizes the free energy for static (equilibrium) configurations at temperature $T$.
For $T\rightarrow 0$, the Fermi-Dirac function reduces to a step function, and only states with positive energy contribute, as in the standard time-dependent BdG approach. 
The Fermi-Dirac weights for $T>0$ provide an approximate way to qualitatively study finite-temperature effects.
This approximation is not controlled--there is no reason to assume that the dynamical thermal effects can be fully encapsulated by the equilibrium Fermi-Dirac distributions--but one expects that this approximation may provide insight into finite-$T$ effects on dynamics close to equilibrium. 
With the exception of one case [Fig.~\ref{fig:kw-spectra}(f)], all simulations discussed here use the $T=0$ formalism.
Further details about the TDASLDA are provided in Appendix~\ref{app:ASLDA-functional}. The implementation is publicly accessible through the W-SLDA toolkit~\cite{WSLDATookit}.
Although very successful in describing the static properties, the TDASLDA still needs to be validated quantitatively for dynamical situations.  
This validation is challenging as typical experimental geometries are too large to be directly simulated by the TDSLDA on current computing platforms.

\begin{figure}[b]
  \includegraphics[width=\columnwidth]{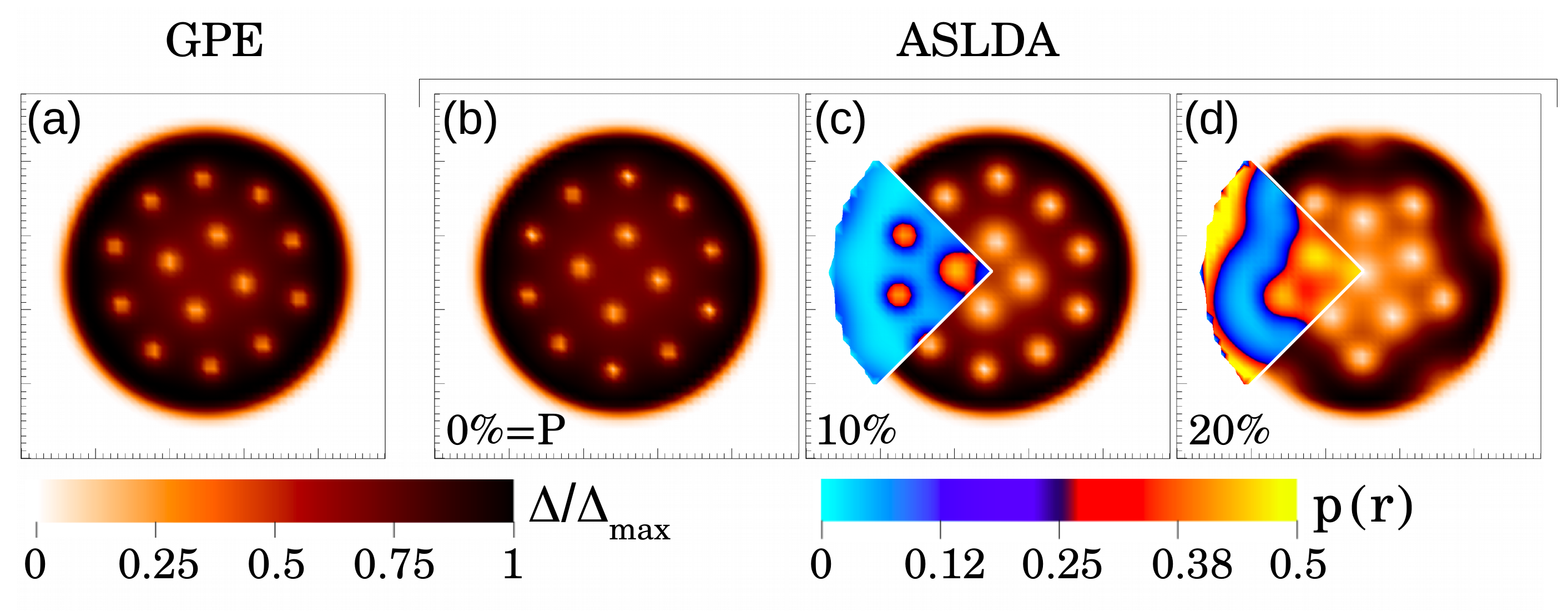}
  \caption{\label{fig:initstates}%
    Cross sections of the initial stationary states used to generate quantum turbulence with the
    (a) \ETF{} and (b-d) ASLDA.
    Panels show the normalized order parameter $\abs{\Delta(\vect{r})}/\Delta_{\max}$ for various spin polarizations: (a)-(b) $P=0$, (c) $P=\SI{10}{\percent}$, and (d) $P=\SI{20}{\percent}$.
    The quarter wedge on the left of (c) and (d) shows the local polarization $p(\vect{r})$.
  }
\end{figure}

We compare this with the \ETF{} as described in Ref.~\cite{ForbesSharma}, modeling the superfluid as a BEC of dimers $(\uparrow\downarrow)$ with Cooper-pair wavefunction $\Psi(\vect{r}, t)$, mass $m_D = 2m$, and dimer density $n_D = n_F/2 = \abs{\Psi}^2$.
The GPE energy density $gn_D^2/2$ is replaced by the UFG energy density $\xiBertsch\mathcal{E}_{FG}(n_F)$:
\begin{gather}\label{eq:ETF}
  \I e^{\I\eta}
  \dot{\Psi}
  = \left(
    \frac{-\hbar^2\nabla^2}{4m} 
    + 2\Bigl(
      \xiBertsch \mathcal{E}_{FG}'(n_F)
    + V - \mu_F
    \Bigr)
    + \Omega\op{L}_{z}
  \right)\Psi. 
\end{gather}
Here, $\mathcal{E}_{FG}' (n_F)$ is the derivative of the energy density of the free Fermi gas $\mathcal{E}_{FG} (n_F)= \frac{3}{5}\varepsilon_F(n_F) n_F$, $\varepsilon_F(n_F) = \hbar^2k_F^2/2m$ is the Fermi energy for a free Fermi gas, $n_F=k_F^3/3\pi^2$ is the total number density of fermions, and $\xiBertsch = 0.373$ is the Bertsch parameter.
The factor of 2 comes from $n_F=2n_D$, i.e.,\@ the dimer chemical potential $\mu_D = 2\mu_F$, etc.
The \ETF{} lacks any mechanism for dissipating hydrodynamic energy into internal degrees of freedom (i.e.\ one-body dissipation).
To account for this, we introduce a small amount of dissipation by hand through the phase $\eta \sim 0.01$.
The presence of a nonzero imaginary time component breaks the conservation of particle number. 
We restore these by dynamically adjusting the chemical potential $\mu_F[\Psi]$. Also, the angular momentum $L_z$ is not conserved, in contrast to TDASLDA calculations. To fix this, the GPE simulations are performed in a rotating frame with constant angular velocity $\Omega$.
It will be shown, that with these modifications, the \ETF{} model~\eqref{eq:ETF} provides a qualitatively similar description of the turbulent phenomena as we see in the fermionic TDASLDA.


We evolve both the TDASLDA and \ETF{} at $T=0$ on the three-dimensional (3D) spatial grids of size $50\times 50\times 100$ with lattice spacing $\dx=\kF^{-1} \approx 0.78\,\healinglength$, where $\healinglength=\kF/\pi\Delta$ is the BCS coherence length.
The gas is confined with a cylindrical trapping potential $V_{\text{ext}}(x^2+y^2)$ with hard walls and a radius $18\,\dx$ and periodic along $z$.
We choose the Fermi momentum of majority spin component $\kF=\sqrt{2\eF} =(6\pi^2n_{\uparrow})^{1/3}\approx 1$, where $n_{\uparrow}$ is the local density of spin-up particles at the center of the trap.
This traps $N_{\uparrow}+N_{\downarrow}\approx \num{3500}$ atoms. For more details related to the simulation process see Appendix~\ref{app:TDASLDA-simulations-details}.

\begin{figure*}[t]
  \includegraphics[width=\textwidth, trim=0 0 0 0, clip]{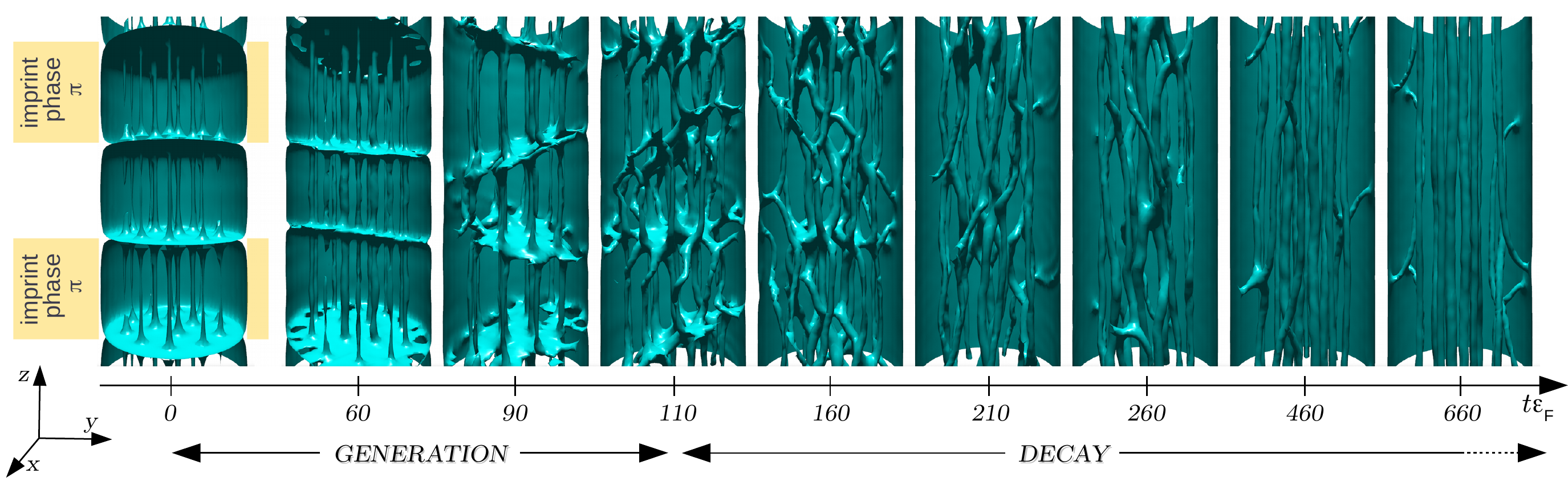}
  \caption{\label{fig:turbulence}%
    Time evolution of spin-balanced ($P=0$) turbulence with the TDASLDA after imprinting four dark solitons at negative times with $\delta \varphi = \pi$ phase shifts across each.
    Frames are isosurfaces of constant order parameter $\abs{\Delta(\vect{r}}$ at various dimensionless times $t\eF$.
    (See the list of movies~\ref{app:List-of-Movies} for movies.) 
    For $t\geq 0$, the system evolves freely, conserving total energy, particle number, and angular momentum $L_z$.
    The two stages of increasing (generation) and decreasing (decay) vortex length are indicated.
  }
\end{figure*}

To generate the quantum turbulence, we start from a self-consistent configuration of $14$ vortices in a lattice that is stationary in a frame rotating with angular velocity $\Omega = -0.05\eF$.
In Fig.~\ref{fig:initstates} we present the solutions for systems with global polarization $P=(N_{\uparrow}-N_{\downarrow})/(N_{\uparrow}+N_{\downarrow}) = \text{\SIlist{0;10;20}{\percent}}$.
The excess spin component accumulates in the vortex cores, changing their size.
The maximum local spin polarization $p(\vect{r})=\bigl(n_{\uparrow}(\vect{r})-n_{\downarrow}(\vect{r})\bigr)/\bigl(n_{\uparrow}(\vect{r})+n_{\downarrow}(\vect{r})\bigr)$ is \SIrange{40}{45}{\percent} in the $P=\SI{10}{\percent}$ case.
Further increase of the spin imbalance alters the lattice structure, as seen already in the case of $P=\SI{20}{\percent}$.
For inhomogeneous systems, as are commonly studied experimentally, physical phase separation seems to dominate, separating superfluid and normal phases, and frustrating the possibility of polarized superfluid states.
This has been observed for rotating spin-imbalanced and strongly-interacting 
Fermi gas~\cite{ZwierleinVortices}, and was confirmed by asymmetric superfluid local density approximation (ASLDA) simulations~\cite{Kopycinski}.
Thus, despite the possibility of polarized superfluid phases, we expect spin imbalance to be a good qualitative probe of the number of unpaired particles, especially when turbulent dynamics create inhomogeneities. For completeness, in Fig.~\ref{fig:initstates}~(a) we display also the initial state obtained by the GPE method. We observe very good agreement when comparing to 
the ASLDA solution for the spin-symmetric system. 

We dynamically perturb this vortex lattice by imprinting four dark solitons in planes approximately perpendicular to the vortices, but with slight tilts to break translation invariance, and imprint a $\pi$ phase shift between each domain.
Due to snaking instabilities, these solitons decay~\cite{MateoYuNian, Cetolietal}, producing new vortices that destabilize the lattice and lead to a vortex tangle.

In Fig.~\ref{fig:turbulence} we show selected snapshots from the TDASLDA simulation for a spin-symmetric system ($P=0$) that demonstrate the full life cycle of the vortex tangle.
This is characterized by two stages: the generation and subsequent decay of the vortex tangle.
During generation, energy from the imprinted solitons is transferred to hydrodynamic flow, increasing the total length $L$ of the vortex lines.
Once the maximum vortex length is reached, the decay process starts and vortex energy is transferred into various internal excitations, including phonons.
This ``decay'' is mediated by vortex dynamics, reconnections, and crossings.
These allow the system to relax back to a simple vortex lattice embedded in an excited superfluid.
After the phase imprinting procedure ($t>0$), the total energy $E_{\textrm{tot}}$, particle number $N$, and $z$ component of the angular momentum $L_z$ are formally conserved.
Numerically, however, the discrete lattice breaks the axial symmetry and $L_z$ is only approximately conserved to an accuracy of about \SI{1}{\percent}.

\section{Rotating quantum turbulence}\label{sec:3}
We analyze the total vortex length $L(t)$ and test the hypothesis that the decay is described by the model
\begin{align}
  \label{eq:dL_dt}
    \frac{dL(t)}{dt} &= -\alpha \bigl(L(t)- L_\infty\bigr)^{1+\epsilon},
\end{align}
where $L_\infty$ is the equilibrium length of the vortices in the lowest-energy state with fixed angular momentum. Details related to the extraction process of vortex lines are described in Appendix~\ref{app:Vortex-detection-algorithm}. 
Vinen turbulence has $\epsilon = 1$~\cite{Vinen:2000}.
For our geometry, $L_\infty = 12L_0$ where $L_0$ is the length of our cylinder, i.e.,\ the ground state contains 12 vortices.
Note that choosing an appropriate value of $L_\infty$ in this model can be problematic as, dynamically, small systems can get stuck in metastable states.
This is demonstrated by the initial states in Fig.~\ref{fig:initstates} which have $L(0)/L_0 = 14$ vortices.

\begin{figure}[ht]
  \includegraphics[width=\columnwidth]{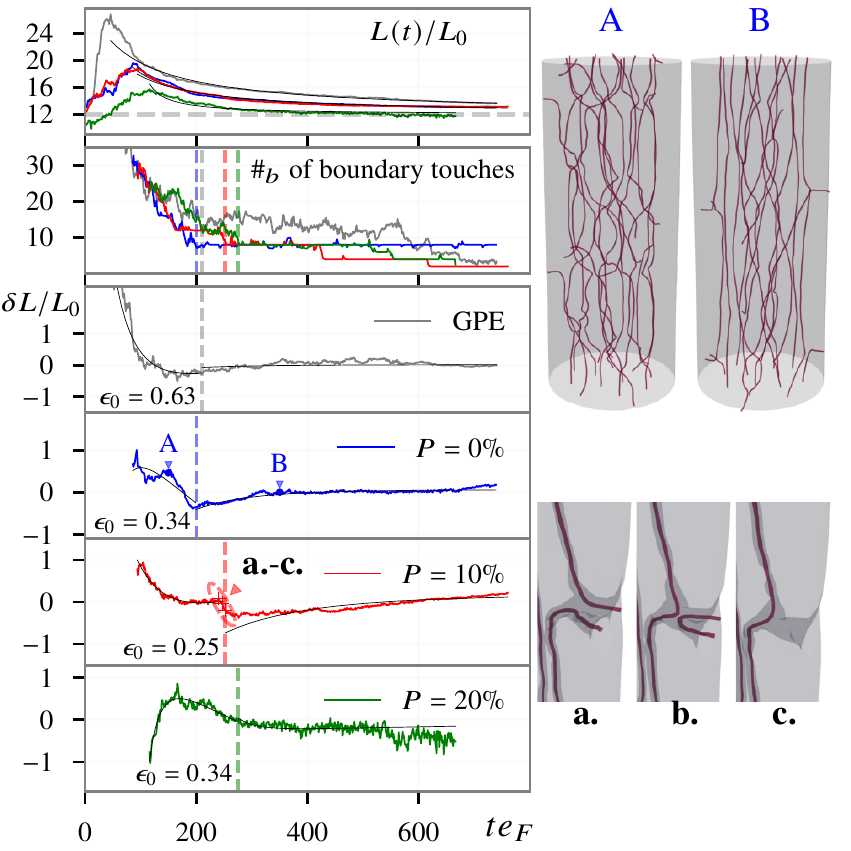}
  \caption{\label{fig:totalL4k}%
    Top panel: Time evolution of the total vortex line length $L(t)/L_0$ normalized to the length of the cylinder with thin curves, $L_{\text{fit}}$, corresponding to best fits with model~\eqref{eq:dL_dt} fixing both $\epsilon=1$ and $L_\infty = 12L_0$.
    Second panel: Number $\#_b(t)$ of vortices touching the boundary of the trap, which decays until time $t_b$ shown as dashed vertical lines, after which it remains relatively constant.
    Bottom four panels: $\delta L(t) / L_0$ where $\delta L = L(t) - L_{\text{fit}}(t)$ for the GPE, and the TDASLDA with spin-imbalance $P\in \{\SI{0}{\percent}, \SI{10}{\percent}, \SI{20}{\percent}\}$ respectively.
    The thin curves in these figures correspond to the second set of fits to model~\eqref{eq:dL_dt} for $t>t_b$ with fixed $L_\infty = 12L_0$ and $\epsilon=1$, and for $t<t_b$ without constraints.
    The best fit exponent $\epsilon_0$ in the latter region is displayed.
    Typical configurations of the unpolarized vortex tangle for two stages of the decay are shown in the top right insets A and B.
    Frames (a - c) on the right correspond to a rapid decrease in vortex length seen for $P=\SI{10}{\percent}$, showing isosurface contours of constant $\abs{\Delta(\vect{r})}$ and thin (red) vortex lines found by our vortex detection algorithm (the algorithm is described in Appendix \ref{app:Vortex-detection-algorithm}).
  }
\end{figure}

Our results are shown in Fig.~\ref{fig:totalL4k}.
The vortex tangle grows until $t_{\max} \approx 100\eF^{-1}$, taking slightly longer with increasing imbalance $P>0$.
We attribute this to the enhanced stability of dark solitons in spin-imbalanced Fermi systems~\cite{SuppressedSolitonicCascade,ReichlMueller,Lombardi2}.
From here, we notice two distinct regimes of decay correlated with $\#_b(t)$, the number of vortices touching the boundary of the trap.
For $t < t_b \approx 250\eF^{-1}$, the number of boundary touches, $\#_b(t)$, 
decreases, while for $t>t_b$ it remains relatively constant.
We thus perform three fits to model~\eqref{eq:dL_dt} (see Appendix~\ref{app:Vortex-length-decay-model} and accompanying code~\cite{figure_code} for details).
We first characterize the size of the fluctuations by calculating a local variance $\sigma^2(t)$ such that $\chi^2(t) \approx 1$ when fitting segments of length $\Delta t \approx 100\eF^{-1}$ with cubic polynomials.
Using these variances, we perform a least-squares fit to the full data for $t>t_{\max}$ fixing $\epsilon = 1$ and $L_\infty = 12L_0$ to obtain a reference $L_\text{fit}(t)$.
The residuals $\delta L(t) = L(t) - L_\text{fit}(t)$, shown in the lower four panels, demonstrate the qualitative change in behavior at $t_b$.
We then individually fit $t<t_b$ with free parameters $\epsilon=\epsilon_0$ and $L_\infty$, and refit $t>t_b$ with fixed $\epsilon = 1$ and $L_\infty=12L_0$.
These fits are shown as thin solid lines in the corresponding residual plots with the best fit value $\epsilon_0$ inset.
This two-component model generally improves the fit, lowering the reduced $\chi^2_r$ by a factor of 2 to 4 compared with the single-component model $L_\text{fit}(t)$.
While the limited size of our system precludes a quantitative extraction of the parameters, it establishes the qualitatively different behavior of these two regimes.

This qualitative behavior seems to be largely insensitive to the spin imbalance of the system.
This suggests that the dynamics in this regime are only weakly affected by the internal structure of vortices, which depends strongly on the spin polarization as seen in Fig.~\ref{fig:initstates}.
The initial decay is accelerated by reconnections close to the boundary that expel vortex segments from the system, which may explain the values of $\epsilon_0$ which differ from that expected for Vinen turbulence~\cite{Mongiovi}.
A particularly large event of this type in the $P=\SI{10}{\percent}$ simulation at $t\approx t_b \approx 250\eF^{-1}$ is shown in the insets Figs. 3(a)-3(c), and is responsible for the localized increase in the decay rate.
Qualitatively similar results were obtained for simulations of rotating \ce{^3He}~\cite{Eltsov1}, where vortex reconnections and annihilations were facilitated by the boundary.
In the second \resprm{(slower)} regime of the decay, reconnection events are rare, taking place mainly in the bulk.
These reconnections generate Kelvin waves that propagate along the vortex lines and dissipate into phonons.

\begin{figure}[t]
  \includegraphics[width=\columnwidth, trim=0 0 0 0, clip]{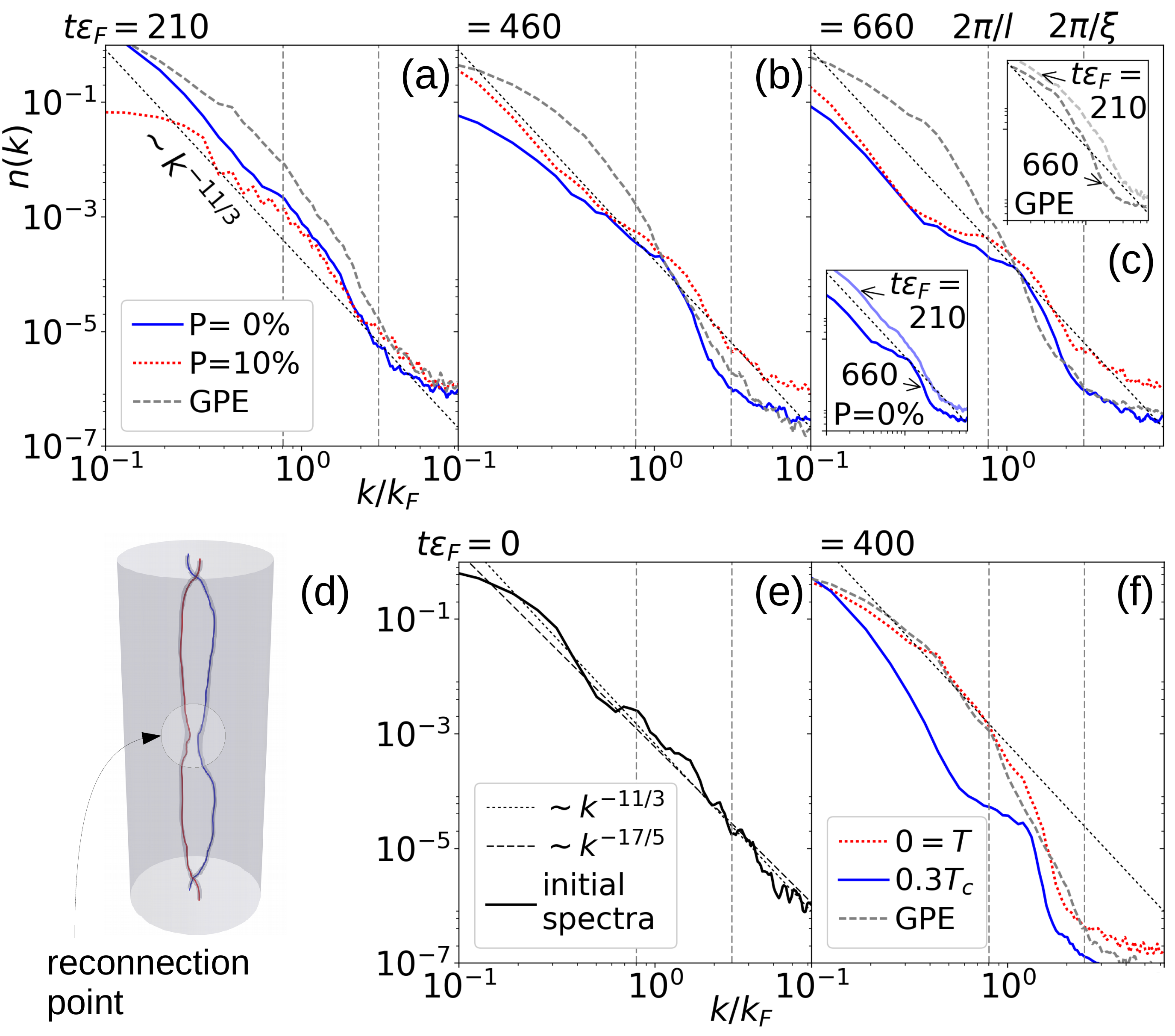}
  \caption{\label{fig:kw-spectra}%
    Kelvin waves spectra $n(k)$ at three times in TDASLDA (spin symmetric and spin polarized) and GPE simulations (a-c). 
    Vertical dashed lines indicate the wave vectors of the two characteristic length scales: the mean intervortex distance ($l$), and the coherence length ($\xi$). 
    The diagonal dotted line indicates the power-law spectrum associated with the wave turbulence~\cite{LvovNazarenko}.
    Insets in (c) show how the spectrum of the GPE (top right) and spin-symmetric TDASLDA (bottom left) change from times $t = 210/\epsilon_F$ to $t = 660/\epsilon_F$. 
    Lower panels show (d) a configuration of two vortex lines about to reconnect, (e )their initial spectra compared with predictions of L'vov and Nazarenko~\cite{LvovNazarenko} (dotted) and Kozik and Svistunov~\cite{KozikSvistunov} (dashed), and (f) the spectra after evolving for $t=400\eF^{-1}$ with the TDASLDA at both zero and finite temperature, and with the GPE.
  }
\end{figure}

Comparing the $P=0$ TDASLDA data with the \ETF{} shows that the latter can be tuned to give the same qualitative features, including the initial generation of turbulence, followed by \respch{two} decay regimes.
To obtain such behavior, however, the dissipation parameter $\eta \approx 0.01-0.02$ must be tuned appropriately to mock up the missing one- and two-body decay mechanisms built into the TDASLDA which convert hydrodynamic energy into internal energy. Also, we maintain the correct value of the angular momentum by simulating in a co-rotating frame with angular velocity $\Omega$ matching the initial state.
The main quantitative difference is that the \ETF{} initially produces more vortices than the TDASLDA, and has a different initial decay exponent $\epsilon_0$.
This deserves further attention, and likely results from the crude mechanism of modeling dissipation with the phase $\eta$.
Nevertheless, the subsequent turbulent decay is quite similar to that of the TDASLDA.

\section{Analysis of Kelvin waves}\label{sec:4}
To provide deeper insight about the decay, we have extracted the averaged spectrum of the Kelvin waves (KWs)
\begin{equation}
  n(k,t)=\frac{1}{N_\text{vor.}(t)}\sum_{l=1}^{N_\text{vor.}(t)}\left(|\tilde{w}_l(k,t)|^2+|\tilde{w}_l(-k,t)|^2\right)
\end{equation}
where $\tilde{w}_i$ is the Fourier transform of the $l$-th vortex parametrized as
$w_l(x,t)=x_l(z,t)+iy_l(z,t)$~\cite{GPE-KW1,GPE-KW2}.
We consider only vortices stretching over the whole cylinder, periodic in $z$, the number $N_\text{vor.}(t)$ of which varies in time. 
Figures~\ref{fig:kw-spectra} 4(a)-4(c) shows the spectra of decaying turbulence for three times.
The first, for $t\lesssim t_b \approx 250\eF^{-1}$, shows a spectrum that approximately follows the dependence derived for a weak KW cascade~\cite{LvovNazarenko,KozikSvistunov} with reconnection events triggering the KW cascade along the vortex lines.
As an example, we show an extracted pair of vortices that are about to reconnect [Fig.4(d)] and the corresponding KW spectrum of these two lines [Fig.4(e)].
This we find as a generic feature, present both in TDASLDA and GPE calculations. Namely, KW spectra of reconnecting vortices is always close to $n(k)\sim k^{-11/3}$ (or $\sim k^{-17/5}$) (see Appendix~\ref{app:Vortex-reconnections-as-trigger-of-Kelvin-wave-cascade} for more details).
We emphasize, as shown in Fig. 4(e), that the accuracy of our data does not allow us to distinguish between the $k^{-11/3}=k^{-3.67}$ spectrum of L'vov and Nazarenko~\cite{LvovNazarenko} and the $k^{-17/5}=k^{-3.4}$ spectrum of Kozik and Svistunov~\cite{KozikSvistunov}. 
In the other panels, we compare spectra only with L'vov \& Nazarenko prediction, since this one is favored by GPE calculations~\cite{GPE-KW1,GPE-KW2}. 
For GPE simulations we observe decay of the spectrum that maintains the overall shape -- see inset in Fig. 4(c), as expected if an energy cascade is present. 
In contrast, the spectrum in the fermionic simulations decays inhomogeneously. In particular, we find that the TDASLDA simulations have an extra suppression for wavelengths with $k<2\pi/l$, where $l$ is the intervortex spacing as compared with the \ETF{} simulations.

To explore the low-$k$ damping, we evolve in isolation from the remaining turbulence and phonons, the two-vortex configuration shown in Fig.~\ref{fig:kw-spectra} (d) for $t\approx 400\eF^{-1}$. We start from a moment where the reconnection event injects energy into the Kelvin waves. As the simulations proceed there are no more reconnection events, which are a necessary ingredient for transferring energy from large to small scales.
Surprisingly, both the TDASLDA and \ETF{} show the same resulting spectrum [Fig. 4(f)] without any long-wavelength suppression.
The spectra decay only for $k\gtrsim 2\pi/l$ where the KW cascade operates.
However, if we repeat these simulations with the TDASLDA at a finite temperature $T=0.3T_c$, a suppression for low-$k$ is seen, similar to that seen in the turbulence scenario from Fig. 4(c). In formulations like vortex filament model or two-fluid Hall-Vinen-Bekarevich-Khalatnikov hydrodynamics, the inclusion of temperature effects is linked with the mutual friction.
This suggests that various excitations (like phonons) in the original $T=0$ simulations of turbulence may act as a thermal reservoir, inducing a more normal component in the TDASLDA that can dampen dynamics at low $k$.
Indeed, we have checked that the amount of energy injected by the phase imprinting procedure is sufficient to heat the gas up to $T=0.5T_c$ (see Fig.~\ref{fig:SM-dyn-potentials} from Appendix~\ref{app:TDASLDA-simulations-details}). 
Similar KW damping is seen in vortex filament models coupled to a normal component (mutual friction), even for weak couplings~\cite{KWBoue,KWHanninen,KWKondaurova}.
The TDASLDA thus provides a fully self-consistent microscopic calculation where a similar phenomenon is observed.
In contrast, the \ETF, which has no normal component, does not show the corresponding damping.

\section{Conclusions}\label{sec:5}
We have simulated dynamical quantum turbulence in a trapped rotating unitary Fermi gas with a microscopic DFT called the TDASLDA, and identified two regimes of decay: one dominated by vortex reconnections near the boundary of the system, and the other dominated by decaying Kelvin waves.
The overall decay pattern can be reproduced by a computationally simpler orbital-free DFT similar to the \ETF\ used to study bosonic superfluids by introducing an appropriately tuned dissipation while manually conserving particle number and angular momentum.
Unlike the \ETF, however, the TDASLDA requires no phenomenological parameters, and, hence, is able to make predictions concerning the turbulent dynamics.
In particular, the results suggest that the TDASLDA accounts for effects related to the presence of a normal component, manifested by damping of the Kelvin waves.
Thus, the TDASLDA may provide a parameter-free self-consistent microscopic theory to study phenomena like mutual friction between superfluid and normal components.

The TDASLDA also allows us to study the effects of spin imbalance, which significantly alters the structure of vortices, but, surprisingly, does not significantly alter the turbulent dynamics.
The procedures used to generate turbulence are compatible with experimental capabilities, paving the way for experiments to address some of the outstanding challenges in the field of quantum turbulence raised in Ref.~\cite{MadeiraReview}.
In particular we have demonstrated a mechanism for generating turbulence in an exotic system with tunable interactions by imprinting solitons that generates turbulence both in the bulk and near the boundaries.
We also demonstrate the inclusion of beyond-mean-field effects through an empirically established DFT that includes induced Hartree term absent in the traditional mean-field BdG approach.
Interestingly, the TDASLDA develops quantum turbulence without the need for any additional dissipation, unlike the \ETF{}, providing insight into the microscopic mechanism for vortex tangle generation discussed in Ref.~\cite{MadeiraReview}.
This provides a means for calibrating simpler \ETF-like theories which may be crucial for understanding quantum turbulence in complex systems such as neutron stars.

While certainly not a complete description of dynamics in the unitary Fermi gas, the TDASLDA seems to incorporate many of the most important ingredients needed to understand fermionic quantum turbulence.  
We hope these results encourage experimentalists to study similar dynamics, providing quantitative validation of the TDASLDA.
This quantitative validation is crucial for using similar theories to study quantum turbulence in neutron stars where experimental validation is impossible. 

\begin{acknowledgments}
  ASLDA calculations were executed by G.W. and K.S., and GPE calculations were executed by K.H. and M.M.F.\@.
  Data analysis was performed by K.K., G.W., K.H., and M.M.F.\@.
  All authors contributed to discussion and interpretation of the results and to writing of the manuscript.
  This work was supported by the Polish National Science Center (NCN) under Contracts No.~UMO-2017/26/E/ST3/00428 (G.W.) and UMO-2017/27/B/ST2/02792 (K.K., P.M.), and the National Science Foundation through Grant No.~PHY-1707691 (K.H., M.M.F.).
  We acknowledge PRACE for awarding us access to the resource Piz Daint based in Switzerland at Swiss National Supercomputing Center (CSCS), Decision No. 2017174125.
  This research used resources of the Oak Ridge Leadership Computing Facility, which is a DOE Office of Science User Facility supported under Contract No. DE-AC05-00OR22725.
  We also acknowledge the Global Scientific Information and Computing Center, Tokyo Institute of Technology, for resources at TSUBAME3.0 (Project ID: hp180066 and hp190063) and the Interdisciplinary Centre for Mathematical and Computational Modelling (ICM) of Warsaw University for computing resources at Okeanos (Grant No.~GA76-13).
\end{acknowledgments}

\appendix
\section{ASLDA functional}\label{app:ASLDA-functional}


%

The structure of the ASLDA functional reads:
\begin{multline}\label{EASLDA}
  \mathcal{E}_\mathrm{\small{ASLDA}} = \alphup(p)\frac{\taup}{2} +  \alphdw(p)\frac{\tadw}{2} + \beta(p)(\nup+\ndw)^{5/3}\\
  + \frac{\gamma(p)}{(\nup+\ndw)^{1/3}}\nu^{*}\nu + \sum_{i=\uparrow,\downarrow}[1-\alpha_{i}(p)]\frac{\bm{j}^2_{i}}{2n_{i}}.
\end{multline}
where normal $n_i$ and anomalous $\nu$ densities are defined via Eqs.~(\ref{eqn:dens_n}) and (\ref{eqn:dens_nu}), while
  \begin{align}
  \tau_{i}(\textbf{r}) &= \sum_{|E_n|<E_c} |\nabla v_{n,i}(\textbf{r})|^2 f_{\beta}(-E_n),\\
    \textbf{j}_{i}(\textbf{r}) &= \sum_{|E_n|<E_c} \textrm{Im}[v_{n,i}(\textbf{r}) \nabla v_{n,i}^{*}(\textbf{r})]f_{\beta}(-E_n),
  \end{align}
stand for the kinetic and current densities, respectively.  
The coupling constants $\alpha_{i}$, $\beta$, and $\gamma$ are polynomial functions of the local polarization of the gas, $p=(\nup-\ndw)/(\nup+\ndw)$.
The polynomial coefficients have been adjusted to quantum Monte Carlo results for both spin-balanced and spin-imbalanced unitary Fermi gas (see Ref.~\cite{LNP__2012} for details related to the fitting procedure).

The term in the functional that depends on the currents $\textbf{j}_{i}$ introduces significant cost to the computation.
This term is responsible for maintaining Galilean invariance of the ASLDA theory.
Since the effective mass was found to be consistent with the bare mass to within $10\%$ for a large range of polarizations, in the calculations we set $\alpha_{i}=1$.
Therefore, the particular form of the functional used in this paper simplifies to:   
\begin{equation}\label{EA1}
  \mathcal{E}_{\alpha=1} = \frac{1}{2}(\taup+\tadw) + \beta(p)(\nup+\ndw)^{5/3} + \frac{\gamma}{(\nup+\ndw)^{1/3}}\nu^{*}\nu.
\end{equation}
The TDASLDA equations are obtained from the condition of extremal action,
\begin{equation}
  S=\int_{t_0}^{t_1} \left( \langle 0(t)| i\frac{d}{dt}| 0(t) \rangle - E(t) \right) dt,
\end{equation} 
with respect to variation of Bogoliubov quasiparticle wave functions, where $|0(t) \rangle$ denotes the quasiparticle vacuum at time $t$ and $E(t)$ is the total energy, 
\begin{equation}
  E(t)=\int \left ( \mathcal{E}_{\small{\alpha=1}}(\bm{r},t) + \sum_{i=\uparrow,\downarrow} V_{i}(\bm{r},t)n_{i}(\bm{r},t) \right ) d\bm{r}.
\end{equation} 
$V_{i}$ is the external potential, described later.
The resulting equations of motion are given by Eq.~(\ref{eqn:TDASLDA}),
where the single-particle Hamiltonian and pairing potentials take, respectively, the following forms:
\begin{align}
  h_{i}(\textbf{r},t) &= -\frac{1}{2}\nabla^2 + \frac{\partial \mathcal{E}_{\small{\alpha=1}}(\bm{r},t)}{\partial n_i}+V_{i}(\bm{r},t)-\mu_i,\label{eqn:sph}\\
  \Delta(\textbf{r},t) &=-\frac{\partial \mathcal{E}_{\small{\alpha=1}}(\bm{r},t)}{\partial\nu^*}. 
\end{align}
Note that the pairing coupling constant $\gamma$ entering the equation for $\Delta(\bm{r},t)$ is the subject of a regularization procedure as described in Ref.~\cite{LNP__2012}.
The chemical potentials $\mu_i$ are meaningful only in the static case of Eq.~(\ref{eqn:TDASLDA}), i.e.\ when $i\frac{\partial}{\partial t}\rightarrow E_n$, where they are adjusted to fix the particle number $N_i$. 
The spin-reversed components of the quasiparticle wavefunctions are obtained via the symmetry relation $u_{n,\uparrow}\rightarrow v_{n,\uparrow}^*$,  $v_{n,\downarrow}\rightarrow u_{n,\downarrow}^*$ and $E_n \rightarrow -E_n$.

For time integration, we used a fifth-order Adams-Bashforth-Moulton (predictor-corrector) scheme.
The integration time step $\Delta t$ was taken to be $\Delta t\,E_{\textrm{max}}=0.035$, where $E_{\textrm{max}}=\pi^2 / 2 {dx}^2$ is estimation of the maximum energy that can be resolved on anemployed spatial lattice with spacing $dx$.

\begin{figure*}[t]
  \includegraphics[width=\textwidth, trim=0 0 0 0, clip]{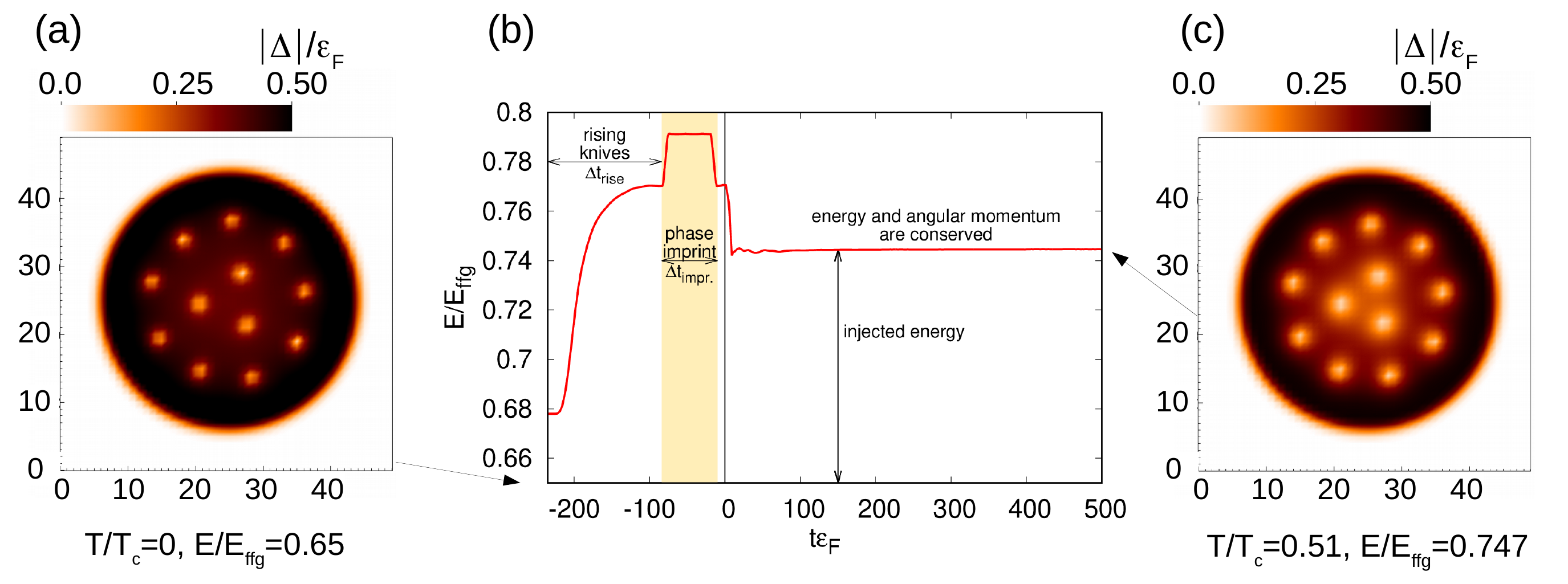}
  \caption{(b): Time evolution of the total energy (expressed in units of free Fermi gas energy $E_{\mathrm{ffg}}=\frac{3}{5}N\eF$) in simulations for $P=0\%$ where turbulence is generated by introducing four knives and the imprinting phase difference is $\pi$. (a) Order parameter distribution of the lowest energy state derived from static calculations at $T = 0$, and (c) the same quantity resulting from the same static calculations but with temperature increased to $T=0.51T_c$.
    Note that the time-dependent runs start from a slightly excited state with 14 rather than 12 vortices and $E\approx 0.68E_{\mathrm{ffg}}$ corresponding to the configuration shown in Fig.~1 of the main text.
  }
\label{fig:SM-dyn-potentials}
\end{figure*}

\section{Simulations details}\label{app:TDASLDA-simulations-details}
\subsection{Generation of the initial state}
The initial state for the simulations is obtained as a solution of static ASLDA equations, which are obtained from Eq.~(\ref{eqn:TDASLDA}) as stationary solutions: 
$v_{i}(\textbf{r},t) = \exp(-i E t)v_{i}(\textbf{r})$,
$u_{i}(\textbf{r},t) = \exp(-i E t)u_{i}(\textbf{r})$.
The external potential has the form (distances are expressed in lattice units $dx$):
\begin{equation}
  \label{eq:Vext}
  V_{\textrm{trap}}(x,y)=
  5\eF\begin{cases}
    0, & \rho \leqslant 17.5,\\
    s(\rho-17.5,7.5), & 17.5<\rho<25,\\
    1& \rho\geqslant 25,
  \end{cases}
\end{equation} 
where $\rho=\sqrt{x^2+y^2}$ is the distance from the symmetry
$z$-axis and $s$ denotes the switching function
\begin{equation}
 s(x,w)=\dfrac{1}{2}+\dfrac{1}{2}\tanh\left[ \tan\left(  \frac{\pi x}{w}-\frac{\pi}{2} \right) \right],
\end{equation}
which smoothly rises from zero to $1$ over distance $w$. Since the trapping potential does not depend on $z$ the quasiparticle wave functions acquire the generic form $\varphi(\bm{r}) \rightarrow \varphi(x,y)e^{ik_z z}$ and the 3D static problem effectively reduces to two-dimensional (2D) problem which needs to be solved for each quantum number $k_z$. 

To start with the solution containing the vortex lattice, the ASLDA equations were solved in the rotating frame by changing the single particle Hamiltonian~(\ref{eqn:sph}):
\begin{equation}
  h_{i}(\textbf{r})\rightarrow h_{i}(\textbf{r})-\Omega L_z,
\end{equation}
where $L_z = -ix\frac{\partial}{\partial y} + iy\frac{\partial}{\partial x}$ is $z$-th component of the angular momentum operator.
The ASLDA equations were solved self-consistently.
Note that the generated state used as the starting point for the time-dependent runs does not correspond to the absolute lowest energy state.
It corresponds to one of excited states, but still stationary.
In Fig.~\ref{fig:SM-dyn-potentials}~(a) we show the lowest energy state under imposed constraints.

\subsection{Dynamical generation of the turbulent state}\label{app:Dynamical-generation-of-the-turbulent-state}
The turbulent state is generated dynamically by applying a time-dependent external potential:
\begin{equation}
  V(\bm{r},t) = V_{\textrm{trap}}(x,y) + V_{\textrm{knives}}(\bm{r},t)+V_{\textrm{impr.}}(\bm{r},t).
\end{equation}
A potential $V_{\textrm{knives}}$ is used to separate the gas into segments (see the first panel of Fig.~2 in the main text):
\begin{equation}
  V_{\textrm{knives}}(\bm{r},t) = \sum_{k=1}^{K}A(t)\exp\left( -\frac{(z-z_k(x,y))^2}{2\sigma^2}\right).
\end{equation}
Here $K$ denotes the number of knives (we did simulations for $K=2$ and $4$) and $z_k$ is the position of a given knife potential:
\begin{equation}
z_k(x,y) = k\frac{N_z}{K} + r_1 x + r_2 y,
\end{equation}
where $r_1$ and $r_2$ are randomly selected small numbers ($r\ll 1$). Thus, the knives are evenly distributed along the $z$ axis with spacing $\frac{N_z}{K}$ and each of them is independently tilted with respect to the $xy$ plane.
The random tilts break periodic symmetry along the $z$ direction. The width of the knife potential was taken as $\sigma\approx 2 dx$.
The time-dependent amplitude of the knife potential was changed in the following way (see Fig.~\ref{fig:SM-dyn-potentials}):
\begin{enumerate}
\item We started simulations with $A=0$ and gradually increased it up to a value $A_{\textrm{max}}=2.5\eF$ in the time interval $\Delta t_{\textrm{rise}}\approx 150\varepsilon^{-1}_{F}$,
\item Next, we kept the amplitude at a fixed value for a time interval $\approx 100\varepsilon^{-1}_{F}$
\item finally, in the short time interval $\Delta t_{\textrm{off}}\approx 10\varepsilon^{-1}_{F}$ we turned off the knives. 
\end{enumerate}
The time when we started to turn off knife potentials is indicated as $t=0$.

The phase imprint is realized by turning on for some time interval $\Delta t_{\textrm{impr}}$ the external potential $V_{\textrm{impr}}$.
This is a constant potential applied  only to selected segments of the cloud, indicated by yellow boxes in Fig.~\ref{fig:turbulence}.
The pairing field is proportional to the anomalous density $\nu \sim \sum_{n}v_{n,\downarrow}^*u_{n,\uparrow}$ and it evolves as $\Delta = e^{2i\mu t}|\Delta|$ where $\mu$ is the chemical potential.
For segments that are the subject of the phase imprinting potential we have $\Delta = e^{2i(\mu - U_0)t}|\Delta|$, where $U_0$ is the strength of the potential.
Consequently, after time $\Delta t_{\textrm{impr.}}$ the phase for imprinted segments gets an extra shift $\delta\varphi=2U_0\Delta t_{\textrm{impr.}}$.
The strengh of the potential $U_0$ is adjusted to introduce the requested phase difference $\delta\varphi$ within a time interval $\Delta t_{\textrm{impr}}\eF\approx 75$.

Figure~\ref{fig:SM-dyn-potentials}(b) is a representative graph showing the total energy change for different stages of the state preparation ($t<0$) and subsequently for the evolution ($t>0$) which is the subject of observation.
The stages related to applying the knives, phase imprinting, and removing knives are clearly visible.
We also checked that the main properties of the tangle evolution do not depend on values of $\Delta t_{\textrm{impr}}$ and $\Delta t_{\textrm{rise}}$. 

Finally, we compare the time-evolved energy of the system to the ground state energy with the same conserved quantities: particle number, polarization, and angular momentum $L_z$.
This difference reflects the amount of energy injected by the preparation procedure, see Fig.~\ref{fig:SM-dyn-potentials}~(b).
We expect that, after sufficiently long evolution, the system will reach an new equilibrium state with the injected energy distributed in phonons and internal degrees of freedom, effectively establishing a sort of thermal equilibrium.
To estimate the effective temperature of this equilibrium state, we perform a static finite-temperature calculation, thermally populating the single-particle states so that the energy matches the total injected energy in our $T=0$ dynamical simulation, again at fixed particle number, polarization, and angular momentum.
To obtain the same energy in a stationary thermodynamic state, we find that we need to set $T=0.51T_c$ [where $|\Delta(0)|/k_B T_c=1.76$; see Fig. 5(c)].
It is currently not known how to precisely relate these two systems---the dynamical equilibrium established in our $T=0$ simulations and the static $T=0.51T_c$ thermodynamic ensemble---but this gives a qualitative estimate of the effective temperatures realized in these simulations.

\subsection{GPE simulations}
To generate the initial state for the GPE simulations, we set our lattice spacing as described in the main text and the chemical potential $\mu_F = \xiBertsch\varepsilon_F$ in the same external potential in Eq.~\eqref{eq:Vext}.
The trap holds $\approx 3700$ particles.
Then we imprint 14 vortices in the same approximate pattern as seen in the ASLDA initial state, and minimize the state with fixed number of particles in a rotating frame with $\Omega = -0.05\varepsilon_F$ to get a solution of the GPE similar to that of ASLDA with $P = 0\%$ in two dimensions.
This solution is then used to populate the 3D state with periodic boundary condition in $z$-direction. The initial state that we generate in this process is similar to the ASLDA initial state, with a couple of small differences in the arrangement of the vortices and number of particles.
These differences do not modify the qualitative physics.

Next, we evolve the 3D state using Eq.~\eqref{eq:ETF} to dynamically generate turbulence following the same procedure described in the previous section.
Then we evolve the turbulent state with different dissipation parameters to study the life cycle of the state. To get the best matching with TDASLDA results we find that $\eta\sim 0.01$ should be used.

\begin{figure}[t]
  \includegraphics[width=\columnwidth, trim=0 0 0 0, clip]{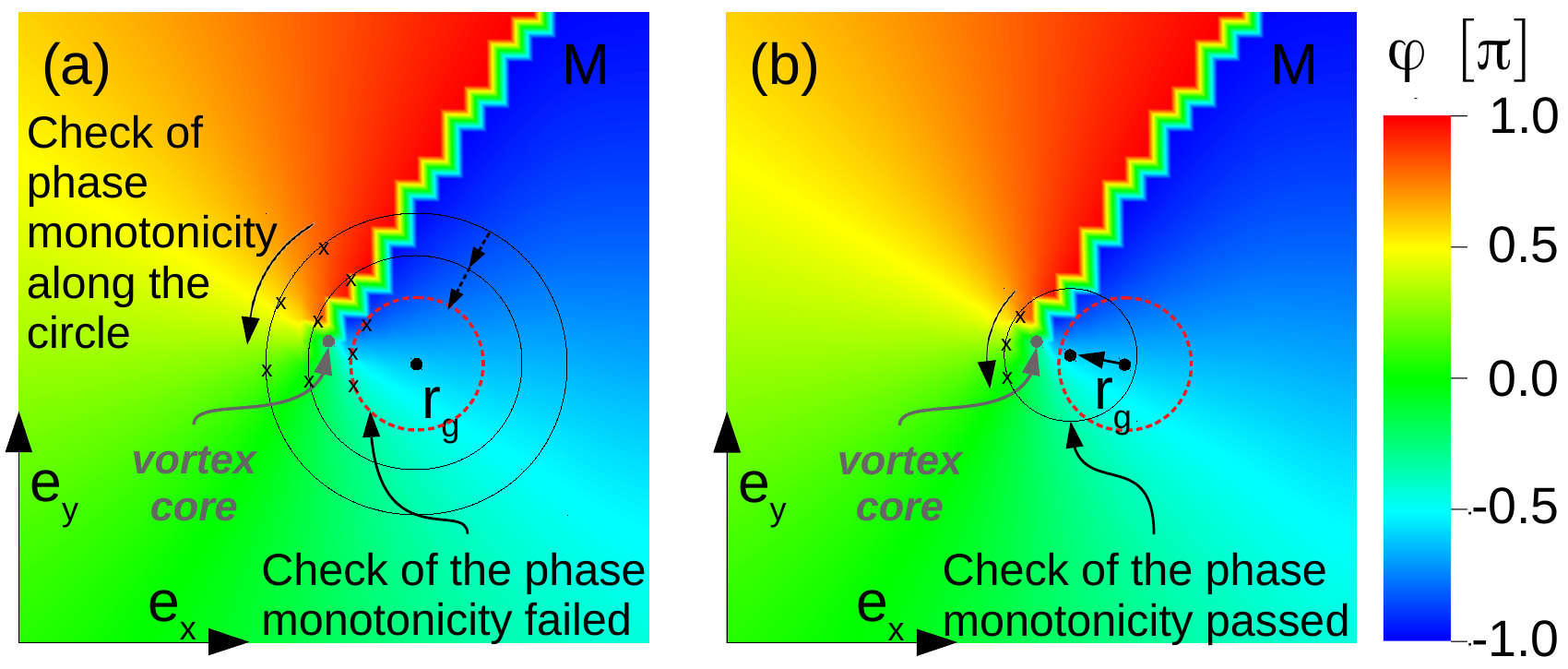}
\caption{Visualization of the main part of the vortex detection algorithm: steps 2(a) and 2(b), see text for more details. The color map encodes the phase pattern of the order parameter $\varphi = \textrm{arg}(\Delta)$. }
\label{fig:detection-algorithm}
\end{figure}


\section{Vortex detection algorithm}\label{app:Vortex-detection-algorithm}
The vortex detection algorithm is based on the modification of the procedure presented in Ref.~\cite{Villoisetal}. The algorithm consists of the following steps:
\begin{enumerate}
\item For a given spatial point $\bm{r}_{g}$ (initial guess), we evaluate the pseudovorticity $\bm{\omega}_{ps}(\bm{r}_g)=\nabla \times \bm{j}(\bm{r})|_{\bm{r}=\bm{r}_g}$ and create a plane $M$ that is normal (perpendicular) to the evaluated $\bm{\omega}_{ps}(\bm{r}_g)$ vector. It is assumed that the guess point is located close to the vortex core. 
\item Within the plane $M$ we search for a point around which the phase of the order parameter $\Delta$ rotates by $2\pi$. The phase is sampled for $k=8$ points uniformly distributed on a circle with radius $R$ constructed around point $\bm{r}_g$. The initial value of the circle radius is $0.75dx$. Next, the vortex core is searched by repeatedly applying the following operations:
\begin{enumerate}
    \item The radius of the circle is decreasing with step $dR=0.05dx$ until the phase pattern along the circle loses the winding property [see Fig.~\ref{fig:detection-algorithm}(a)]. 
    \item The position of the circle is shifting $\bm{r}_g+d\bm{r}\rightarrow \bm{r}_g$, in such a way to restore the expected phase pattern [see Fig.~\ref{fig:detection-algorithm}(b)].
    \end{enumerate}
    Steps (a) and (b) are executed until the radius reaches the value $R=0.05dx$ which sets the accuracy of the vortex location detection. The center of the circle is assigned to the vortex position $\bm{r}_v$.
\item For the vortex core position $\bm{r}_v$ the pseudovorticity is  recalculated $\bm{\omega}_{ps}(\bm{r}_v)=\nabla \times \bm{j}(\bm{r})|_{\bm{r}=\bm{r}_v}$ and a new guess point is created: $\bm{r}_g = \bm{r}_v + \frac{dx}{3}\frac{\bm{\omega}_{ps}(\bm{r}_v)}{|\bm{\omega}_{ps}(\bm{r}_v)|}$.
\end{enumerate}
Steps  $1-3$ are executed until the reconstructed vortex line either touches boundaries or creates a closed loop. The first guess point for each line is located by analyzing the phase pattern for lattice plates.

The procedure described above was used to obtain vortex line positions for TDASLDA and GPE simulations after every $\delta t \varepsilon_F=1$. Such a time resolution guarantees that in the co-rotating frame the ending of each vortex line lies sufficiently close (we assumed distance of $1.5 dx$) to the same vortex line ending in the previous time step, which allows for temporal tracking of vortices. We use abrupt changes of vortex end positions between two consecutive time frames as indicator of reconnection, annihilation, or tearing of vortices.

Having extracted vortex lines, the following quantities were analyzed:
\begin{itemize}
\item The total length of vortices, $L(t)$:  This is calculated as a sum of distances between consecutive points lying on distinct vortex lines. (The points are given by the detection algorithm.)
\item Number of ends of vortices touching the boundary, $\#_b(t)$: This is extracted as vortex ends lying sufficiently close ($2 dx$) to the isosurface of low density at the boundary of the system.
\item Number of vortices, $N_V(t)$: The vortex lines can be distinguished based on the mutual distance between pairs of the points given by the detection algorithm with the periodic boundary condition along the$z$ axis being taken into account.
\end{itemize}

\begin{figure*}[!h]
  \includegraphics[width=\textwidth]{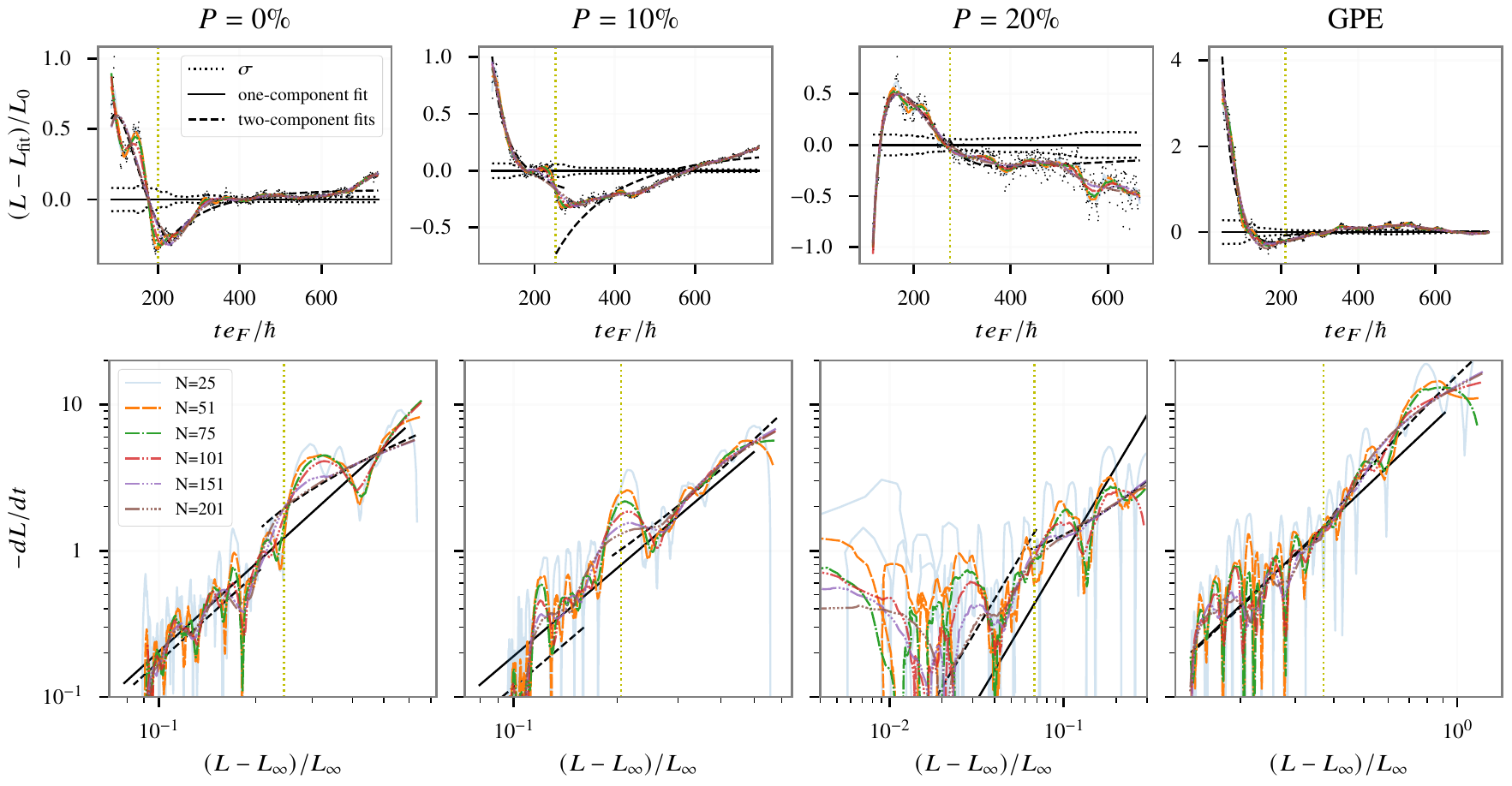}
  \caption{Modeling the decay of the vortex length $L(t)$.
  Top row: Vortex length $\Delta L(t)/L_0$ as in Fig.~3 of the main text with the standard deviations $\sigma(t)$ used as weights in the fits.
  Bottom row: $-dL(t)/dt$ as a function of $(L-L_\infty)/L_\infty$.
    This should be linear if model (1) from the main text is satisfied.    The solid black line is a one-component best fit to the entire set of data $t>t_{\max}$ with $\epsilon = 1$ (Vinen turbulence) and $L_\infty = 12L_0$ [ground state shown in Fig.~\ref{fig:SM-dyn-potentials} (a)].
  The black dashed lines are the best two-component fits to the regions $t<t_b$ with free $\epsilon_0$ and $L_\infty$ as parameters, and then for $t>t_b$ with constrained $\epsilon = 1$ and $L_\infty = 12L_0$.
  The colored curves correspond to numerically computed derivatives of the actual data after smoothing with a Savitzky-Golay filter of length $N$.
  We compute $\sigma(t)$ by performing a least-squares fit of the data in segments of length $N=101$ to a third-order polynomial, and adjust the standard deviation $\sigma(t)$ (dotted lines in the upper panels) over each interval so that $\chi^2 = 1$ for these fits.
  This gives an estimate for the size of the numerical fluctuations.
  The choice of window size is not unique, but $N=101$ does a reasonable job of smoothing over dynamical features, while still preserving the dominant long-term trends.
  Our results change little with $N \in \{51, 75, 101, 151\}$.
  To explore this analysis and for code to reproduce this figure, please see Ref.~\cite{figure_code}.
  }
\label{fig:derivatives}
\end{figure*}

\section{Vortex length decay model}\label{app:Vortex-length-decay-model}
We present the details of our analysis of the vortex length data. To describe the two regimes of decay we use a two-component fit using the model presented in the main text. In order to compute the derivatives of the data we needed to perform smoothing of the data. In Fig.~\ref{fig:derivatives}, we present the results obtained from using different window sizes for the smoothing filter, as there is no obvious choice. We used a window length which preserves the dominant long-term trends.

\begin{figure*}[htbp]
  \includegraphics[width=\columnwidth, trim=0 0 0 100, clip]{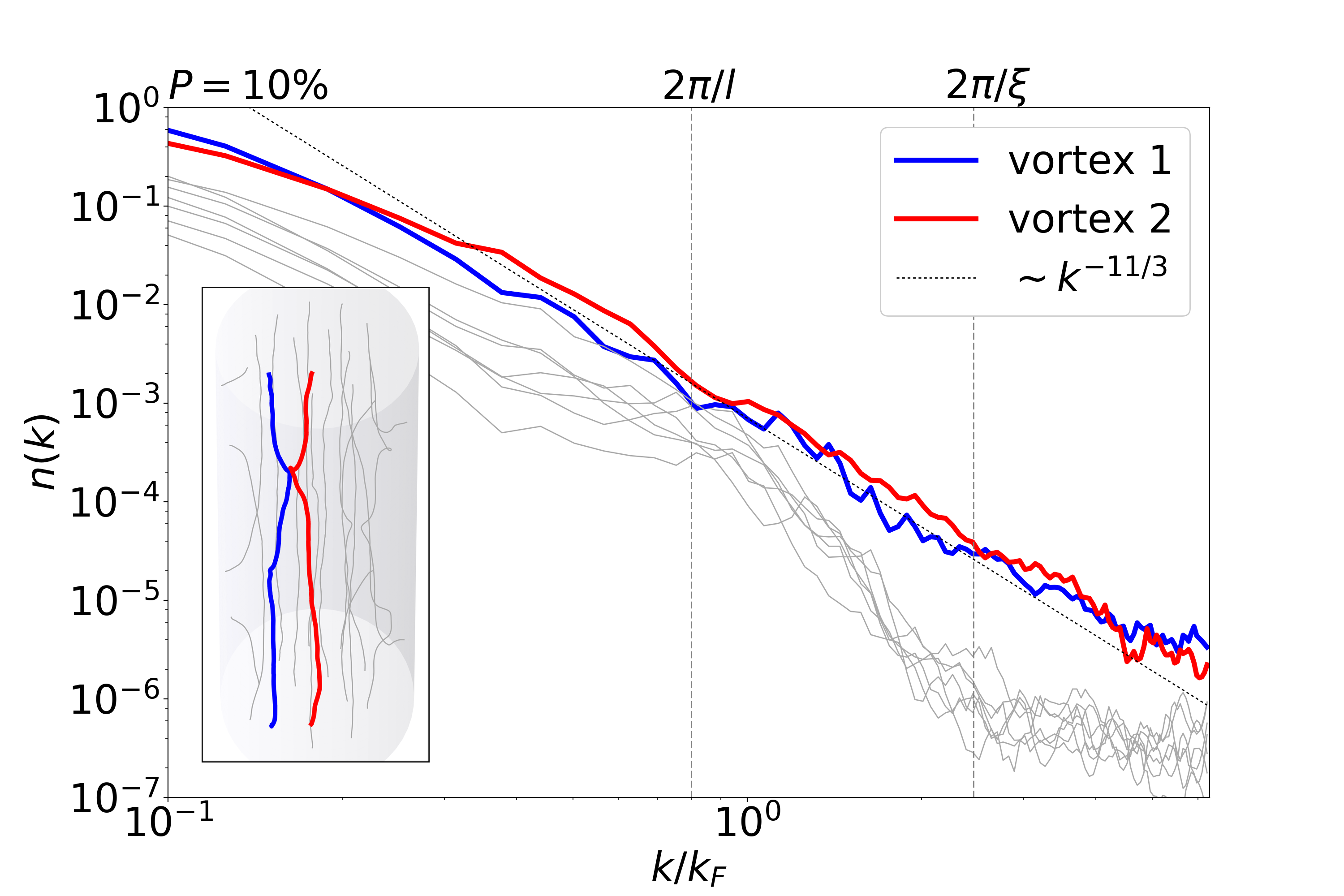}
  \includegraphics[width=\columnwidth, trim=0 0 0 100, clip]{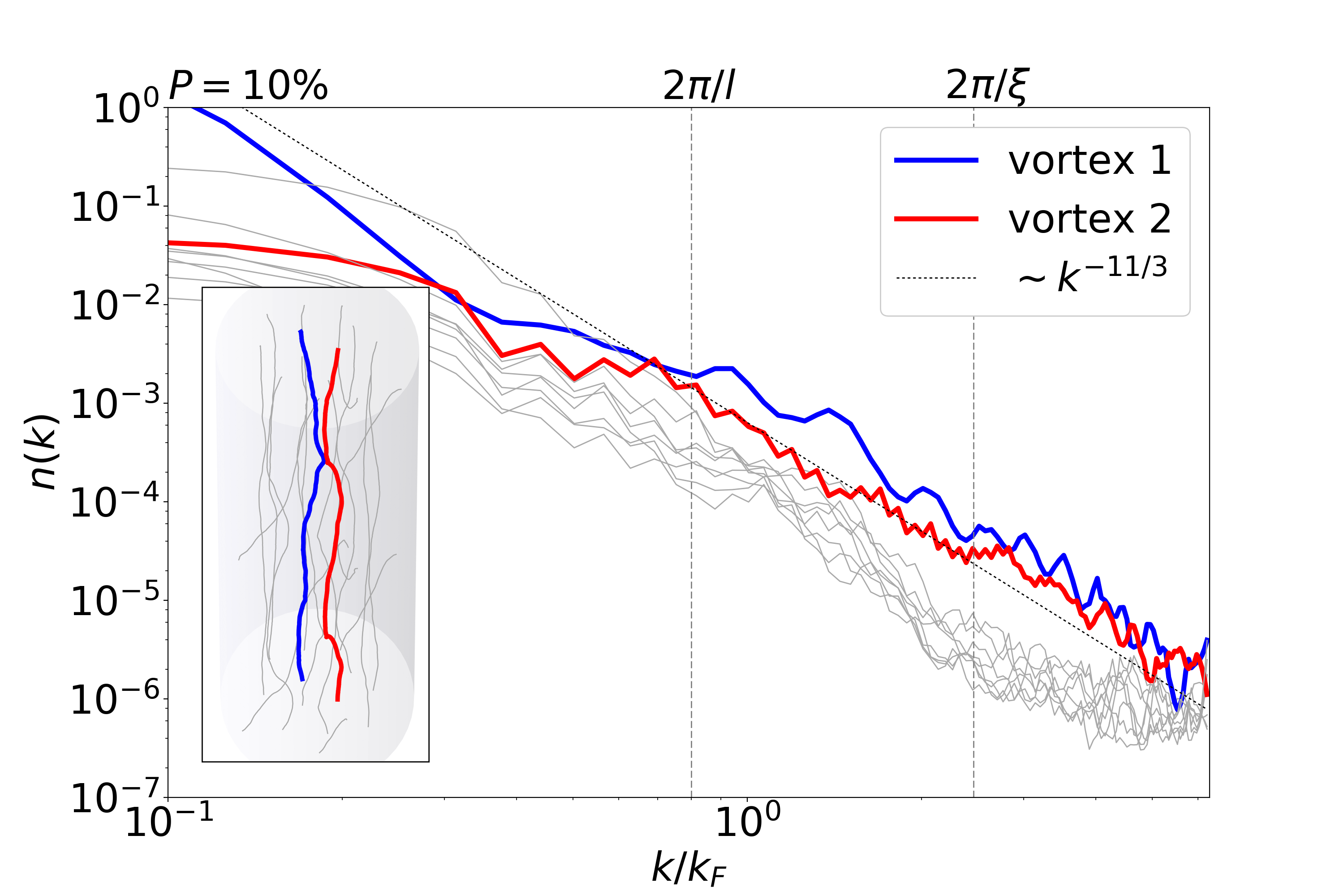}
\caption{KW spectra of individual vortices for two selected moments in time, corresponding to the vortex reconnection event. Vortices that undergo the reconnections are marked by (bold) red and blue lines. The results are obtained by means of the TDASLDA method for spin-symmetric (left) and spin-imbalanced (right) systems. As the generic feature, we find that the reconnections populate vortices with the spectra which is compatible with KW cascade spectra, marked by the dotted line. In insets, we show corresponding vortex line configurations, with highlighted vortices that undergo the reconnection. 
}
\label{fig:slda-vr}
\end{figure*}
\section{Vortex reconnections as trigger of Kelvin-wave cascade}\label{app:Vortex-reconnections-as-trigger-of-Kelvin-wave-cascade}
The analysis of Kelvin waves clearly demonstrates that emergence of the spectra compatible with the wave turbulence cascade $n(k)\sim k^{-11/3}$ (or $\sim k^{-17/5}$) is due to vortex reconnections. Namely, we find as a generic feature (present both in Superfluid local density approximation and GPE calculations) that each reconnection event populates the crossing lines with the corresponding spectra. In Fig.~\ref{fig:slda-vr} we show spectra of each individual vortex lines for two selected moments in time, where the reconnection takes place. It is clearly seen that the reconnecting vortices are attributed with spectra compatible with the KW spectra. Note that, in Figs.~\ref{fig:kw-spectra}(a)-~\ref{fig:kw-spectra}(c) we present averaged spectra over all lines.
\section{List of Movies}\label{app:List-of-Movies}
Below we provide the list of accompanying movies. All movies are also accessible on YouTube. There are two types of visualization:
\begin{description}
 \item[vis1] Visualization of contour plot of $|\Delta|$. Left upper panel shows corresponding total length of vortex lines, while left bottom panel shows integrated along $z$ direction value of $|\Delta|^2$, i.e. $\rho_{\textrm{2d}}(x,y) = \int |\Delta(x,y,z)|^2\,dz$.
  \item[vis2] Visualization of vortex tangle. Central panel shows vortex lines given as output of vortex detection algorithm and isocontours of $\vert \Delta\vert$, left upper panel shows corresponding total length of vortex lines, left bottom panel shows positions at which reconnection events detected (current ones are marked with red color) and right upper and bottom panels are number of vortex ends touching boundary $\#_b$ and number of vortices $N_V$, respectively.
\end{description}
List of movies with description:
\begin{description}
\item[Movies 1 and 7] TDASLDA and \ETF{} simulation demonstrating development and decay of rotating turbulence in the spin-symmetric UFG. The turbulence was generated by imprinting 4 solitons with a phase difference of $\pi$. Frames from this movie were used for Fig.~2 of main paper.\\[2mm]
             \textbf{TDSLDA}:\\
             Files: \verb|Movie1-vis1|, \verb|Movie1-vis2|\\
             YouTube: \url{https://youtu.be/5NiFRymHPYk},\\
    \phantom{YouTube: }\url{https://youtu.be/yAtXReicxBY};\\
             \textbf{\ETF} (in a co-rotating frame):\\
             Files: \verb|Movie7-vis1|, \verb|Movie7-vis2|\\
             YouTube: \url{https://youtu.be/EFwJ2vOtmzQ},\\
    \phantom{YouTube: }\url{https://youtu.be/tPS1WIF7rUU}

\item[Movie 2] TDASLDA simulation demonstrating development and decay of rotating turbulence in the spin-imbalanced UFG with total polarization $P=10\%$. All other conditions are the same as for \textit{Movie~1}.\\[2mm]
             Files: \verb|Movie2-vis1|, \verb|Movie2-vis2|\\
             YouTube: \url{https://youtu.be/imnsogKVhHQ},\\
    \phantom{YouTube: }\url{https://youtu.be/ng44ojYDuHw}

\item[Movie 3] The same as \textit{Movie~2}, but for total polarization $P=20\%$\\[2mm]
             Files: \verb|Movie3-vis1|, \verb|Movie3-vis2|\\
             YouTube: \url{https://youtu.be/h82ApFUwkI8},\\
    \phantom{YouTube: }\url{https://youtu.be/6eEPsNJrJ58}

\item[Movies 4 and 8] TDASLDA and \ETF{} simulations where turbulence was generated by imprinting 2 solitons with phase difference of $\pi$. Qualitatively the movie is similar to \textit{Movies~1 and 7}.\\[2mm]
             \textbf{TDSLDA}:\\
             Files: \verb|Movie4-vis1|, \verb|Movie4-vis2|\\
             YouTube: \url{https://youtu.be/zNilo_bSrCk},\\
    \phantom{YouTube: }\url{https://youtu.be/v72nPV70MxQ};\\
             \textbf{\ETF} (in a co-rotating frame):\\
             Files: \verb|Movie8-vis1|, \verb|Movie8-vis2|\\
             YouTube: \url{https://youtu.be/rrNRMMkqTj4},\\
    \phantom{YouTube: }\url{https://youtu.be/SeYT8Evynjo}

\item[Movies 5 and 9] The same as \textit{Movies~4 and 8}, except the value of the imprinted of phase difference which here is $\pi/2$. This imprint generates moving solitons which induce amplification of Kelvin waves.\\[2mm]
             \textbf{TDSLDA}:\\
             Files: \verb|Movie5-vis1|, \verb|Movie5-vis2|\\
             YouTube: \url{https://youtu.be/U8r5yiz3PPY},\\
    \phantom{YouTube: }\url{https://youtu.be/EAzhKumyddI};\\
             \textbf{\ETF} (in a co-rotating frame):\\
             Files: \verb|Movie9-vis1|, \verb|Movie9-vis2|\\
             YouTube: \url{https://youtu.be/-FSJkcwZGOM},\\
    \phantom{YouTube: }\url{https://youtu.be/gREn5kB3qOE}

\item[Movie 6] Movie showing result of collision of clouds with vortices where the phase imprint procedure is not applied. In such a case the vortex tangle is not generated.\\[2mm]
             Files: \verb|Movie6-vis1|, \verb|Movie6-vis2|\\
             YouTube: \url{https://youtu.be/HYSou4qK8fQ},\\
    \phantom{YouTube: }\url{https://youtu.be/S3WTbvIFqY4}

\end{description}

\vspace{1\baselineskip} 

\providecommand{\citenamefont}[1]{#1}
\providecommand{\bibnamefont}[1]{#1}
\providecommand{\bibfnamefont}[1]{#1}
\providecommand{\bibinfo}[1]{}
\providecommand{\bibfield}[1]{}
\providecommand{\Eprint}[2]{}
\providecommand{\BibitemOpen}{}
\providecommand \bibitemNoStop [0]{.\EOS\space}%
\providecommand \EOS [0]{\spacefactor3000\relax}%
\providecommand \BibitemShut  [1]{\csname bibitem#1\endcsname}%

\end{document}